\documentclass[%
 reprint,
longbibliography,
 amsmath,amssymb,
aps,
prl,
]{revtex4-1}
\setcitestyle{numbers,square}

\usepackage{braket}
\usepackage{graphicx}% Include figure files
\usepackage{dcolumn}% Align table columns on decimal point
\usepackage{bm}% bold math
\usepackage[normalem]{ulem} % strikethrough
\usepackage{color}

\newcommand{\orcid}[1]{\href{https://orcid.org/#1}{\includegraphics[width=8pt]{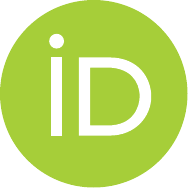}}}

\usepackage{hyperref}% add hypertext capabilities
\hypersetup{
    colorlinks,%
    citecolor=blue,%
    linkcolor=blue,%
    urlcolor=blue
}

\begin{document}

\title{Emergent quasiparticles in Euclidean tilings}

\author{F. Crasto de Lima \orcid{0000-0002-2937-2620}} 
\email{felipe.lima@lnnano.cnpem.br}
%\affiliation{Brazilian Nanotechnology National Laboratory, CNPEM, \\ C.P. 6192, 13083-970, Campinas, SP, Brazil}

\author{A. Fazzio \orcid{0000-0001-5384-7676}}
\email{adalberto.fazzio@lnnano.cnpem.br}
\affiliation{Brazilian Nanotechnology National Laboratory CNPEM, \\ C.P. 6192, 13083-970, Campinas, SP, Brazil}

\date{\today}

\begin{abstract}
Material's geometrical structure is a fundamental part of their properties. The honeycomb geometry of graphene is responsible for the arising of its Dirac cone, while the kagome and Lieb lattice hosts flat bands and pseudospin-1 Dirac dispersion. These features seem to be particular for few 2D systems rather than a common occurrence. Given this correlation between structure and properties, exploring new geometries can lead to unexplored states and phenomena. Kepler is the pioneer of the mathematical tiling theory, describing ways of filing the euclidean plane with geometrical forms in its book {\it Harmonices Mundi}. In this letter, we characterize $1255$ lattices composed of the euclidean plane's k-uniform tiling, with its intrinsic properties unveiled — this class of arranged tiles present high-degeneracy points, exotic quasiparticles, and flat bands as a common feature. Here, we present aid for experimental interpretation and prediction of new 2D systems.
\end{abstract}

\maketitle

\begin{figure*}
\includegraphics[width=2.05\columnwidth]{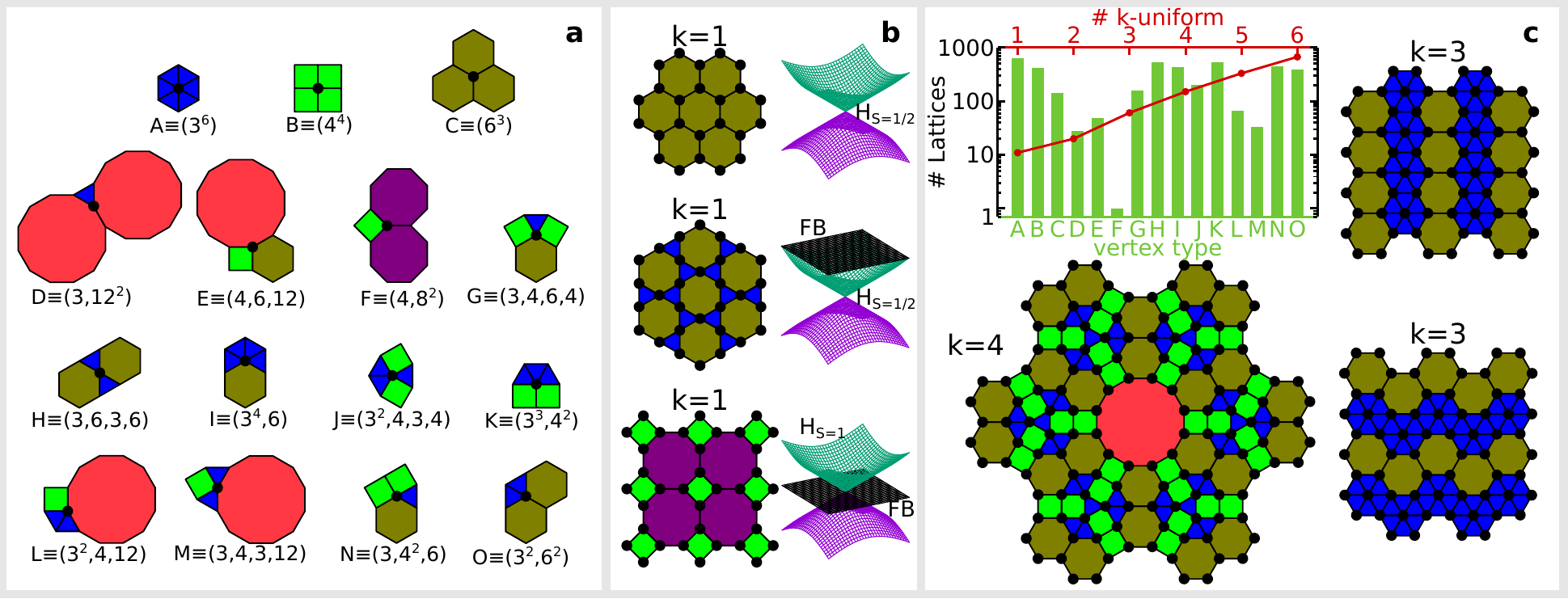}
\caption{\label{lattices} \textbf{Lattices with k non-equivalent vertex.} \textbf{a.} All possible vertex for translational invariant uniform Euclidean tilings, labeled according to conventional notation \cite{MMgrunbaum1977}. \textbf{b.} Examples of k=1 uniform tiling and arising quasiparticles. \textbf{c.} Graph showing the number of lattices with k non-equivalent vertexes (red) and the number of lattices with at least one vertex of type (green) indicated in \textbf{a}, together with three tiling examples.}
\end{figure*}

One question that arises is how to describe all the possible perfectly flat 2D materials? Commonly, we consider real materials formed by entities (orbitals, closed boundaries, waveguides) pinpointing a lattice site, while lines can connect those sites indicating a pair interaction (for instance, as chemical bonds). A precise way to represent such lattices is to tile the Euclidean plane with regular polygons. Such an assumption allow us to describe most of the observed 2D systems, from translational periodic \cite{NATREVCHEMjin2017} to quasicrystal structures \cite{NATUREbursill1985}. Their composing types of vertexes can classify the tiling. There are only 21 types of vertexes (encounter of three or more polygons) composed of regular polygons. Considering translational periodicity, 15 from the 21 possible vertexes can tile the plane \cite{MMgrunbaum1977}, which are shown in Fig.~\ref{lattices}\textbf{a}.
 
An infinite number of lattices arise combining different vertexes. We can classify such lattices concerning the type of vertexes they are composed, and if these sites are equivalent by crystal symmetry operations. A arrange of tiles composed of regular polygons containing k non-equivalent vertexes is denoted as k-uniform tiling. Interestingly, the most popular synthesized 2D materials are described by the k-uniform tiling. Besides the graphene honeycomb and the kagome lattice, we clarify in Fig.~\ref{lattices}\textbf{c} that k=3 tilings describe 2D borophene phases. Exotic fermions can emerge in this systems, for instance, as shown in Fig.~\ref{lattices}\textbf{b}, the spin-1/2 Dirac cone of graphene, the flat band of the kagome lattice, and the pseudospin-1 Dirac of the square-octagonal lattice. The characterization of all possible k-uniform tiling is still an open mathematical problem. However, up to k=6, all possible tilings have been found, together with a few k=7, totaling 1255 classified k-uniform tilings \cite{saesa}.

%%%%%%FIG

Besides monoatomic materials exploration \cite{AMglavin2020} and its natural discovery, rational design approaches allow precise systems. Currently, we are facing combined efforts between theoretical and experimental groups to realize new 2D materials. For instance, in rational designed metal-organic frameworks, that allow for a tiling based filing of the plane \cite{NATCOMMliang2019}. Additionally, by varying molecules and annealing process, many structures have been constructed in molecular self-assembly \cite{NATCOMMcheng2018, ACSNANOfeng2019, NATCOMMkormos2020}. Scanning Tunneling Microscopy (STM) designed surface crystals allows higher control over the lattice formation \cite{NATNANOpolini2013, NATCOMMcollins2017, NATPHYslot2017, NLcrasto2020}, expanding the capability of lattices exploration. Such high control can also be found in photonic crystals \cite{NATjoannopoulos1997}, vibroacoustic materials \cite{MTming2009, NATPHYSmeeussen2020}, and circuit lattices \cite{NATkollar2019}. Exploring the beam shape techniques such lattices can also be constructed with highly ordered structures in cold-atoms \cite{NATNANOpolini2013}, and projecting superlattices directly over 2D materials \cite{PRRkim2020}.

A similar problem arises when describing dispersive eigenstates in periodic (i) classical system, for instance, in elastic media or gyroscopic lattices (mass lumped-parameter system), or even (ii) circuit lumped-parameter systems; (iii) electronic crystals, (iv) photonic crystals or (v) cold-atoms in optical lattices. We can describe those systems as quasiparticles moving in periodic lattices. Ultimately we are dealing with an eigenvalue problem of localized elements coupled through some terms. In this letter, we have characterized the energy dispersion of the arising quasiparticles. We consider each vertex of the tiling structure representing a localized entity (orbital, mass, atom, wave) while each edge represents the nearest-neighbor (NN) interaction between these entities. Therefore, a similar hamiltonian structure will be applicable in each particular case. Taking the electronic system as a case of study, we can write
\begin{equation}
H = \sum_{ij} (1-\delta_{ij})\, t_{ij} \, c_{i}^{\dagger} c_{j},
\end{equation}
where $c_{i}$/$c_{i}^{\dagger}$ are the annihilation/creation operators of a electron in the ith site; $(1-\delta_{ij})$ sets all on-site energies to zero; and $t_{ij}$ describes the rate of $i \rightarrow j$ site transition, i.e. the hoping strength of an electron going from ith to the jth site. If we focus on planar isotropic entities, i.e., $s$, $p_z$, $d_{z^2}$ orbitals for the electronic case, $t_{ij}$ assume a simpler distance dependence between ith and jth site $t_{ij} = t_{ij}(d_{ij})$. We can parametrize such interaction concerning the NN distance ($d_{nn}$), $t_{ij} = t \exp \left[ -\alpha (d_{ij} - d_{nn}) \right]$, in such a way that if $\alpha \gg d_{nn}^{-1}$ only NN hoping is significative, while for $\alpha \sim d_{nn}^{-1}$ further neighbors interaction become relevant. Such representation has shown to capture the essence of many materials, for instance, the graphene Dirac cone \cite{RMPcastro2009}, MOF kagome structures \cite{PRBcrasto2017}, and other 2D systems \cite{PCCPcrasto2019}. In the reported results we have taken $t=1$, $d_{nn}=1$, $\alpha\rightarrow \infty$ for the NN hoping only system, and $\alpha=5/d_{nn}$ (unless otherwise stated) for the long-range (LR) coupling. The detailed lattices structure and band dispersion are present in the supplementary information. Here, it is worth emphasizing that LR coupling englobes more than the next-nearest neighbor (NNN) hopping, and taking $\alpha=5/d_{nn}$ leads to the reasonable value for the NNN hoping fraction of graphene $\exp \left[-5(d_{nnn} - d_{nn})/d_{nn} \right] \sim 0.03$ \cite{RMPcastro2009}.

%%%%%%FIG
\begin{figure*}
\includegraphics[width=1.5\columnwidth]{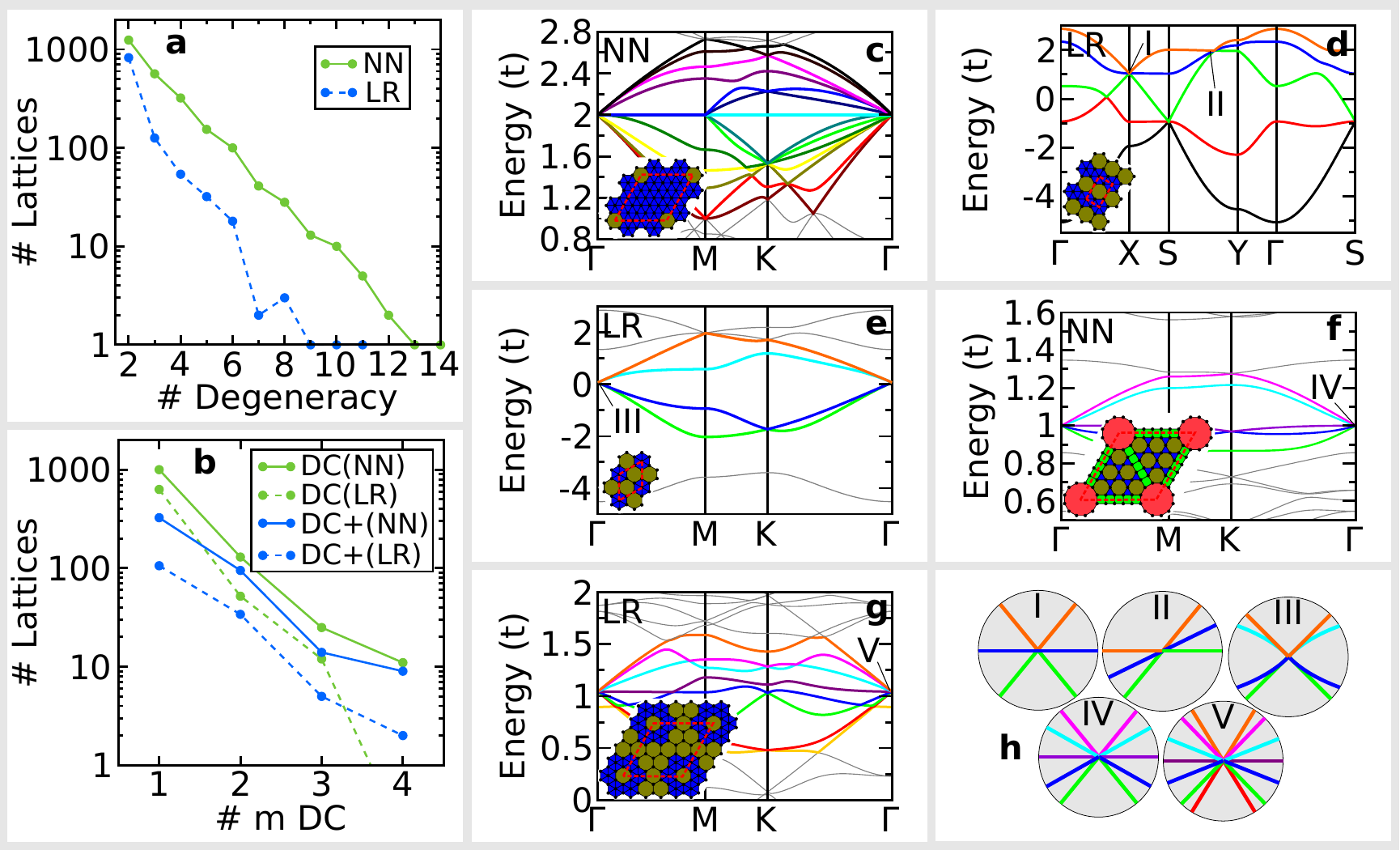}
\caption{\label{deg} \textbf{Degeneracy points and Dirac cones.} \textbf{a.} Number of lattices with at least one point with a given degeneracy order. \textbf{b.} Number of lattices with $m$ degenerated Dirac cones (DC) and $m$ Dirac cones further degenerated with another non-linear band (DC+). Example of lattices hosting \textbf{c.} a 14-fold, \textbf{d.} two 3-fold, \textbf{e.} a 4-fold, \textbf{f.} a 5-fold, and \textbf{g.} a 7-fold degeneracy point. \textbf{h.} Zoom in the marked degeneracies of the example lattices.}
\end{figure*}

An astonishing feature arising in these k-uniform tiling is high-degeneracy points. Spacegroup symmetries alone (wallpaper group in 2D case) can predict only 3-fold (2-fold) degeneracies in spinless systems. However, here we found up to 14-fold degeneracy points. Such points predicted here, rather than being accidental \cite{NATMAThuang2011}, are protected by site permutation symmetries \cite{PRBcrasto2020}. Higher degeneracy points are usually associated with emergent phenomena in materials. For example, supercollimation \cite{PRBfang2016}, Klein tunneling \cite{PRBurban2011}, Weyl points \cite{NATrao2019}, and from an applied point of view with enhanced thermoelectricity \cite{AEMfu2014}. Although the appealing emergent phenomena, few systems have been shown to provide high-degeneracy points, particularly for higher-pseudospin Dirac effective quasiparticles. Here, 519 lattices host at least one 3-fold degeneracy point from which 119 are robust against long-range coupling, as shown in Fig.~\ref{deg}{\bf a}. We can see an almost exponential decaying of the number of lattices hosting a given degeneracy order within the k-uniform tiling for $k \leq 7$. Including long-range coupling can break degeneracies; however, a considerable number of lattices retain such points. For instance, we find 18 lattices hosting 6-fold, while one lattice host a 9-, 10- and 11-fold robust degeneracy.

%%%%%%FIG
\begin{figure*}
\includegraphics[width=1.6\columnwidth]{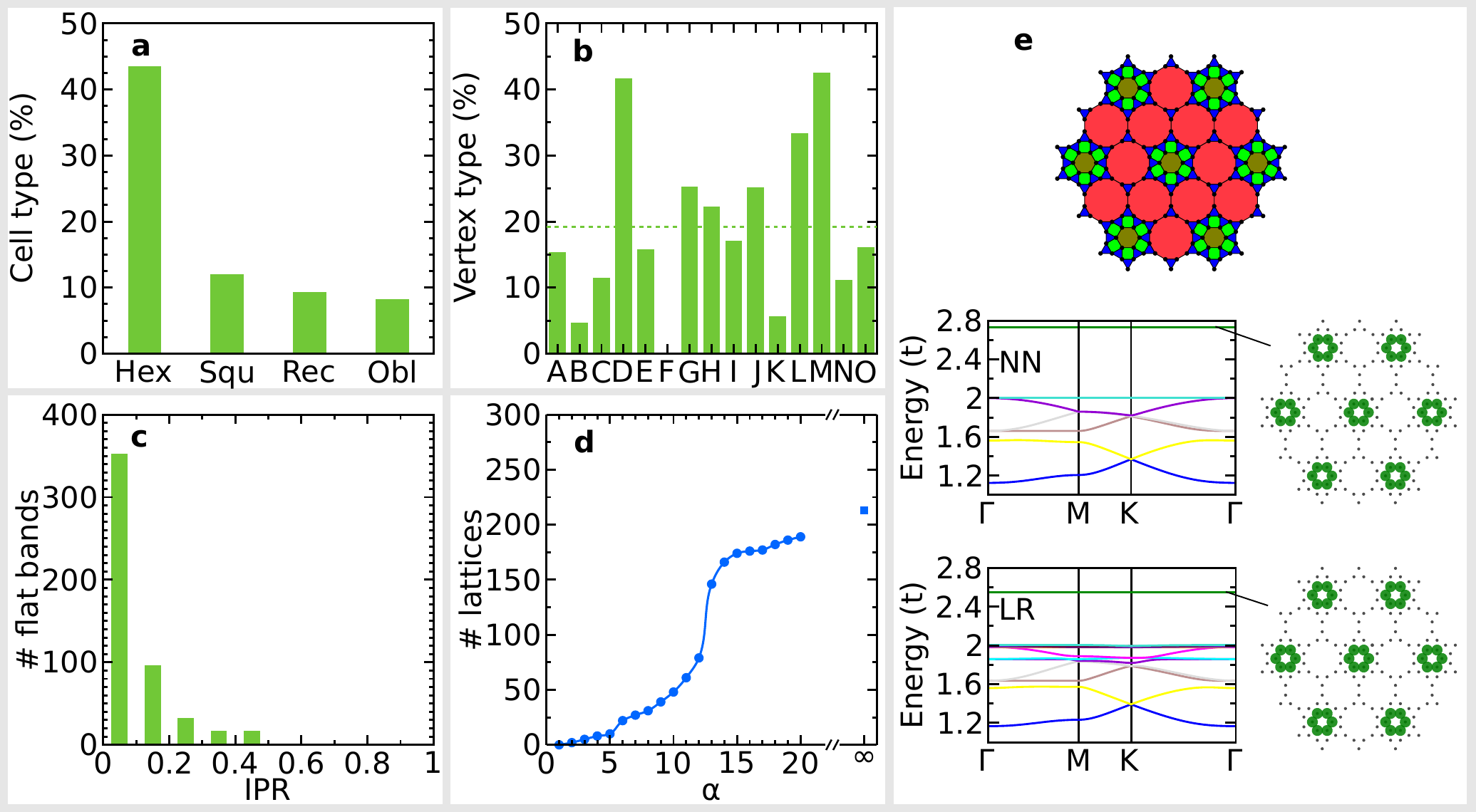}
\caption{\label{flat} \textbf{Flat bands in k-uniform tiling.} \textbf{a.} Percentage of hexagonal, square, rectangular, and oblique lattices with at least one flat band. \textbf{b.} Percentage of vertex type of lattices with at least one flat band. \textbf{c.} Inverse participation rate of the flat bands. In \textbf{a}-\textbf{c} the results are with NN coupling. \textbf{d.} The observed number of lattices with flat bands as a function of the long-range factor $\alpha$. \textbf{e.} k=4 lattice hosting a flat band robust against long-range coupling.}
\end{figure*}

Remarkably, these degeneracy points can lead to new quasiparticles in condensed matter systems that do not have high-energy counterparts. For instance, the pseudospin-S Dirac fermions described by higher-order Dirac equation $H_S = \vec{\sigma}_{S} \cdot {\vec{k}}$, where $\vec{\sigma}_S$ denotes the spin $S$ angular momentum matrix. For $S=1/2$, $\vec{\sigma}_{1/2}$ becomes the Pauli matrix-vector, and the Hamiltonian describes the Dirac fermion present in graphene. In general, for a half-integer spin ($S=(2n+1)/2$), such an equation describes a set of $m=n+1$ degenerated Dirac cones. In contrast, for integer spin ($S=m$), the set $m$ of Dirac cones further degenerate with an additional flat band. In Fig.~\ref{deg}{\bf c} we see that the found 14-fold point is composed of eight linear dispersive bands (four Dirac cones), five flat bands and, one quadratically dispersive band along the $\Gamma$--M direction. Among the studied lattices, we have classified the number of lattices hosting at least $m$ Dirac cones (DC), or $m$ Dirac cones degenerated with at least one flat-band or non-linear dispersive band (DC+). As shown in Fig~\ref{deg}{\bf b}, similar exponential decaying of the number of lattices with the increase of degeneracy order is seeing.

The arising degeneracy points have a rich diversity in their dispersions.  In Fig.~\ref{deg}\textbf{d} we shown a lattice with a pseudospin-1 Dirac quasiparticle [Fig.~\ref{deg}\textbf{h} panel I], but also a 3-fold type-II like Dirac dispersion [Fig.~\ref{deg}\textbf{h} panel II]. This lattice is one of the 106 lattices hosting an LR robust point formed by Dirac cone degenerated with an additional band (DC+ with m=1 in Fig.~\ref{deg}\textbf{b}). It is worth pointing out that no correlation between the vertex types of the lattices and the high-degeneracy points are present. Rather than being local dependent, such degeneracies are a collective effect of the lattice sites. For instance, as shown in the lattice of Fig.~\ref{deg}\textbf{f}, composed of 5 different types of vertexes, a 5-fold pseudospin-2 Dirac [Fig.~\ref{deg}\textbf{h} panel IV] arises. Focusing on vertex composed of triangles and hexagons [A, C, H, I, and O of Fig.~\ref{lattices}\textbf{a}] which leads to triangular based systems, as the ones constructed in controlled STM atomic positioning \cite{NLcrasto2020}, different quasiparticles can be designed. Besides the 14-fold and 3-fold degeneracy of Figs.\ref{deg}\textbf{c} and \textbf{d}, in Fig.\ref{deg}\textbf{e} and \textbf{f} we highlight a 4-fold pseudospin-3/2 Dirac dispersion [Fig.~\ref{deg}\textbf{h} panel III] and a 7-fold pseudospin-3 Dirac quasiparticle [Fig.~\ref{deg}\textbf{h} panel V], respectively. The common appearance of such degenerated Dirac cones in the k-uniform tiling opens opportunities to conetronics manipulations \cite{2DMwu2016}.

Together with the Dirac quasiparticle's attention in graphene, we see a grown interest in electronic flat bands.  They have emerged as an alternative for high-temperature fractional quantum Hall phase without an external magnetic field \cite{PRLneupert2011}. The flatness of the band defines an energy scale to which other effects can compete. For a perfect flat band, any weak perturbation can have a relevant energy contribution. As a consequence, flat bands arising in the condensed matter have shown to present unconventional magnetism \cite{NATPHYSyin2019}, superconductivity \cite{NATcao2018, NATPHYbalents2020}, and a slow light behavior \cite{NATPHObaba2008}.

A given band was considered flat when its dispersion (bandwidth) were lower than $10^{-4}\,t$. For NN coupling, perfect flat bands arise (null bandwidth), for instance, as in the kagome case Fig.~\ref{lattices}\textbf{b}. Indeed, for NN coupling, 213 lattices hosting at least one flat band are present among the studied tilings. Those flat bands occur mostly within hexagonal systems, as shown in Fig~\ref{flat}\textbf{a}. Here, $43\%$ of all hexagonal tilings have at least one flat band. The same is true for around $10\%$ in other cells. The correlation between geometry and the presence of flat bands can also be observed among the different lattice vertexes, Fig.~\ref{flat}\textbf{b}. More than $40\%$ of lattices with at least one D or M vertex [Fig.~\ref{lattices}\textbf{a}] and more than $30\%$ with L vertex present flat bands. Indeed, comparing the average vertex percentage, the green dashed line of Fig.~\ref{flat}\textbf{b}, indicates that some vertex types have a stronger correlation with the appearance of flat bands.

To show the delocalization nature of the flat band, we calculate the inverse participation rate (IPR)
\begin{equation}
P^{-1} = \left( N \sum_i | \Psi_i |^4 - 1 \right)/\left( N-1 \right)
\end{equation}
with, $\Psi_i$ the orbital projected wave function and N the lattice's number of sites. The inversion participation rate is bounded to $0 \leq P^{-1} \leq 1$, being $P^{-1} = 1$ a fully localized state, and $P^{-1} = 0$ a fully delocalized state. In Fig.~\ref{flat}\textbf{c} we shown the number of flat band states within all the 1255 tilings with $P^{-1}$ within intervals of $0.1$. Our results show that these flat bands are delocalized, instead of coming from lone pairs, with most of such states with $P^{-1}<0.1$.
 
The picture presented above of structural correlation and delocalized IPR is preserved considering LR couplings. However, LR interaction couples flat band states from neighboring cells leading to a dispersive state. In Fig.~\ref{flat}\textbf{d} the number of flat bands for a given LR parameter $\alpha$ is presented. Up to $\alpha = 14/d_{nn}$ more than 150 lattices preserves the flat band character. The transition observed around $\alpha = 13/d_{nn}$ can be understood as $70 \%$ of the lattices have the next-nearest neighbor (NNN) distance being $d_{nnn}=\sqrt{2}\,d_{nn}$, while $30\%$ with $d_{nnn} = \sqrt{3}\,d_{nn}$. Therefore in such a point, the shorter $d_{nnn}$ lattices have significant LR interaction while the longer $d_{nnn}$ is still two orders of magnitude lower. Impressively, increasing such coupling range to $\alpha = 5/d_{nn}$, ten lattices still host a flat band. As shown in Fig~\,\ref{flat}\textbf{e} for one of those lattices, the preserved flat band arises from orbitals further separated from its periodic images. Therefore an even higher coupling range is required for the flatness disappearance.

The controlled experimental realization of different lattices has recently grown in photonic, STM molecular positioning, cold-atoms, and other systems. They allow a rich playground for the exploration of new phenomena in exotic dispersive bands. Here we have shown that, rather than being a rare feature, such states vastly occur among the k-uniform tilings. Indeed, we have predicted here higher pseudospin Dirac points, until now hidden from possible realizations. Our theoretical approach can be broadly interpreted within different systems. The explored systems unveil a class of systems to which we give a guide for experimental lattice construction and data interpretation. 

\section*{Acknowledgments}
The authors acknowledge financial support from the Brazilian agencies FAPESP (grants 19/20857-0 and 17/02317-2).

\bibliography{bib} %Produces the bibliography via BibTeX.

\end{document}